\documentclass[twocolumn,superscriptaddress,amsmath,amssymb,showpacs,prb]{revtex4-2}

\usepackage{graphicx}
\usepackage{color}
\renewcommand{\phi}{\varphi}

\begin{document}

\title{Thermodynamics of spin crossover in ferropericlase: an improved LDA+$U_{sc}$ calculation}

\author{Yang Sun}
	\affiliation{Department of Applied Physics and Applied Mathematics, Columbia University, New York, NY 10027, U.S.A.}
\author{Jingyi Zhuang}
	\affiliation{Department of Earth and Environmental Sciences, Columbia University, New York, NY 10027, U.S.A.}
	\affiliation{Lamont–Doherty Earth Observatory, Columbia University, Palisades, NY 10964, U.S.A.}

\author{Renata M. Wentzcovitch}
    \email [Email: ]{rmw2150@columbia.edu}
	\affiliation{Department of Applied Physics and Applied Mathematics, Columbia University, New York, NY 10027, U.S.A.}
	\affiliation{Department of Earth and Environmental Sciences, Columbia University, New York, NY 10027, U.S.A.}
	\affiliation{Lamont–Doherty Earth Observatory, Columbia University, Palisades, NY 10964, U.S.A.}

\begin{abstract}

We present LDA+$U_{sc}$ calculations of high-spin (HS) and low-spin (LS) states in ferropericlase (fp) with an iron concentration of 18.75$\%$. The Hubbard parameter $U$ is determined self-consistently with structures optimized at arbitrary pressures. We confirm a strong dependence of $U$ on the pressure and spin state. Static calculations confirm that the antiferromagnetic configuration is more stable than the ferromagnetic one in the HS state, consistent with low-temperature measurements. Phonon calculations guarantee the dynamical stability of HS and LS states throughout the pressure range of the Earth mantle. Compression curves for HS and LS states agree well with experiments. Using a non-ideal mixing model for the HS to LS states solid solution, we obtain a crossover starting at $\sim$45 GPa at room temperature and considerably broader than previous results. The spin-crossover phase diagram is calculated, including vibrational, magnetic, electronic, and non-ideal HS-LS entropic contributions. Our results suggest the mixed-spin state predominates in fp in most of the lower mantle.

\end{abstract}

\date{Feb. 20, 2022}

\maketitle

\section{Introduction}

Ferropericlase (fp) is the second most abundant mineral in the Earth's lower mantle. It may be responsible for up to $\sim$20 vol$\%$ of this region \cite{a1}. It is a solid solution (Mg$_{1-x}$Fe$_x$)O of MgO and FeO in the rocksalt-type (B1) crystal structure, with $x_{Fe}$=0.15-0.20 in the lower mantle. Its high-pressure electronic properties, spin state, and phase stability are critical to understanding the properties of and processes taking place in the mantle. In particular, iron in fp undergoes a pressure-induced spin-crossover from a high spin (HS) state with S=2 to a low spin (LS) state with S=0. This spin-crossover has attracted extensive research interest because it can have critical geophysical consequences, e.g., a density increase \cite{1}, a bulk modulus softening \cite{2}, thermoelastic anomalies  \cite{3,4}, etc.

Experiments have reported an HS-LS crossover pressure in the range of 40-70 GPa at room or lower temperatures \cite{5,6,7,8,9}, with $x_{Fe}$ around 0.20. High temperature leads to an increase in the spin-crossover pressure range and in the crossover onset pressure. This is caused by a mixed HS-LS state (MS) \cite{1,6,10} caused entropic contributions. Because of the strongly correlated nature of the 3$d$ electrons in Fe and the large supercells used to study the fp solid-solution by first-principles, the HS-LS crossover diagram has been challenging. Tsuchiya $et\ al.$ \cite{1} first performed LDA+$U$ calculations to study the HS and LS crossover with HS/LS configuration entropy and magnetic entropy. Later Ref.\cite{3} extended LDA+$U$ calculations to include vibrational effects based on a virtual-crystal model. The model was also used to calculate thermodynamic anomalies in fp \cite{2}. While LDA+$U$ can reasonably address the electronic structure of the correlated 3d electrons of iron, its performance on the crossover pressure highly depends on the $U$ value \cite{11,12,13,14,15,16}. It has been argued that the complete dependence of $U$ on pressure/volume, structure, spin state, or even pseudopotential should be taken into account if one is to make predictions of phase transitions at extreme conditions. In this work, we employ a recent implementation of the self-consistent calculation of the Hubbard parameter based on density-functional perturbation theory (DFPT) \cite{17}. Using LDA+$U_{sc}$ calculations, we compute the spin crossover diagram for fp with $x_{Fe}$=0.1875. This calculation differs from previous ones from our group by computing the HS and LS states' vibrational spectra vs. volume and going beyond the ideal solid-solution model by computing the excess free energy due to HS-LS interaction in the MS state. 

In the next section, we describe the computational details of the first-principle calculations. In Sec. III, we first analyze the electronic structure of ferropericlase with $x_{Fe}$=0.03125. In Sec. IV, we present static calculations of the spin-crossover of fp with $x_{Fe}$=0.1875 at T=0K. In Sec. V, we compute the phonon spectra and consider the finite temperature effect on the spin crossover by including various entropic contributions and non-ideal mixing effects. We summarize all findings in Sec. VI.

\section{Methods}

LDA+$U$ calculations were performed using the simplified formulation of Dudarev \textit{et al.} \cite{18} as implemented in the Quantum ESPRESSO code \cite{19,20}. The local density approximation (LDA) was used for the exchange-correlation functional with spin polarization. The projector-augmented wave (PAW) data sets from the high-accuracy version of PSlibrary \cite{21} were employed with valence electronic configurations 3s$^2$3p$^6$3d$^6$4s$^2$, 2s$^2$2p$^6$3s$^2$, and 2s$^2$2p$^4$ for Fe, Mg, and O, respectively. A kinetic-energy cutoff of 100 Ry for wave functions and 600 Ry for spin-charge density and potentials were used. In all cases, atomic orbitals were used to construct occupation matrices and projectors in the LDA+$U$ scheme. A cubic supercell of B1 structure with 64 atoms was constructed for the current study, i.e., (Fe$_x$Mg$_{1-x}$)$_{32}$O$_{32}$. The $2\times2\times2$ $k$-point mesh was used for Brillouin zone integration. Structure optimization was performed by relaxing atom positions with a force convergence threshold of 0.01 eV/$\text{\AA}$. The convergence threshold of all self-consistent field (SCF) calculations was $1\times{10}^{-9}$ Ry.

\begin{figure}[b]
\includegraphics[width=0.47\textwidth]{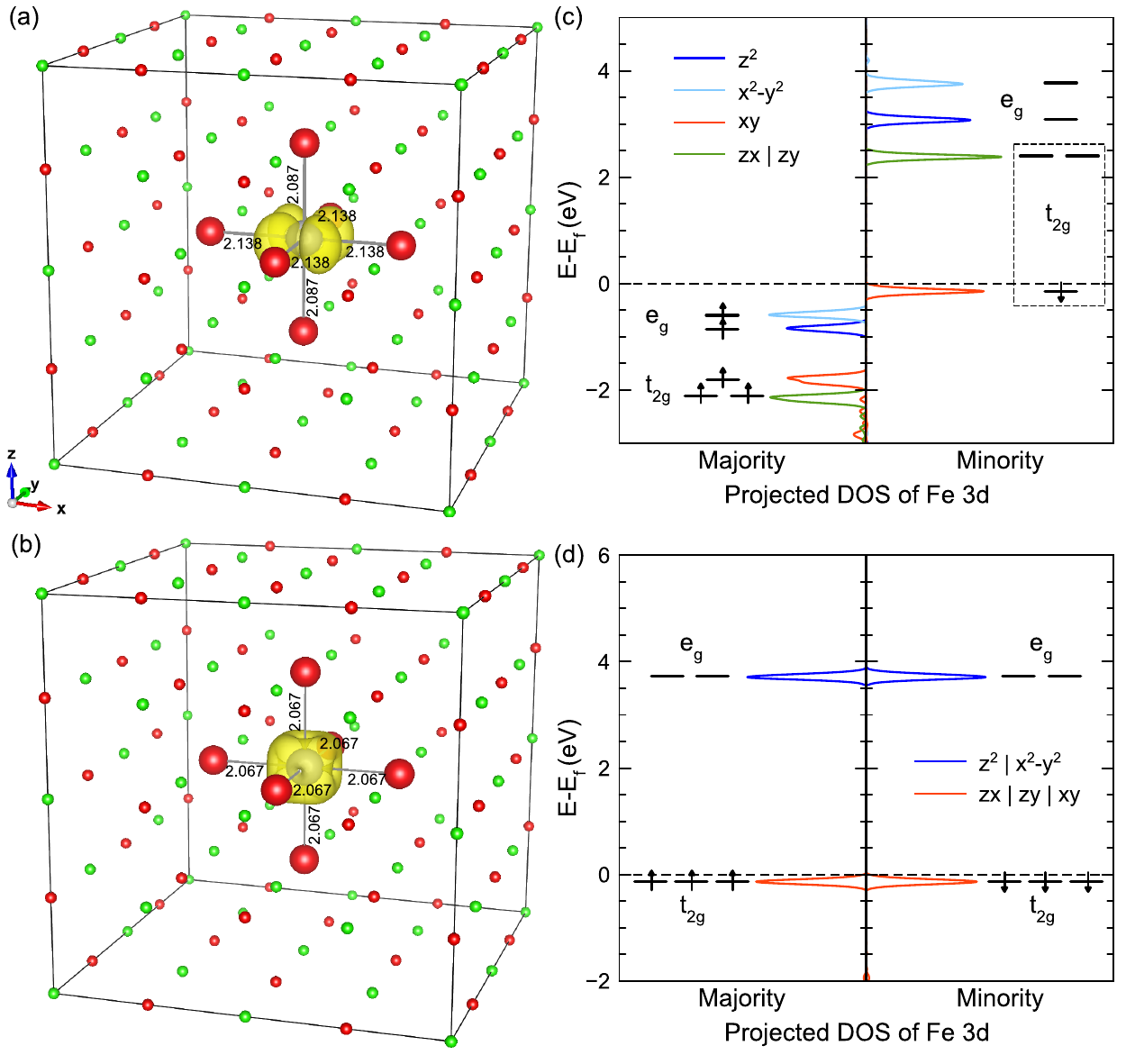}
\caption{\label{fig:fig1}(a) The HS state of fp3 at P=0GPa. Gray is Fe, green is Mg and red is O. The charge density (yellow) is shown for the Fe minority electron occupying the d$_{xy}$ orbital. T Fe-O bond lengths in angstrom ($\AA$) are shown inside the octahedron. Mg and O without Fe-O bonds are drawn smaller for clarity. (b) The LS state of fp3 at P=0GPa. The charge density (yellow) is shown for the occupied t2g orbitals. (c) and (d) Projected density of state (DOS) for Fe 3d orbitals in HS and LS fp3 at P=0GPa. The schematic shows the energy splitting of e$_g$ doublet and t$_{2g}$ triplet.}
\end{figure}

The Hubbard correction \cite{22} was applied to Fe-3$d$ states. The Hubbard parameter $U$ was computed using density-functional perturbation theory \cite{17} implemented in the Quantum ESPRESSO code. The convergence threshold for the response function is $1\times{10}^{-6}$ Ry. An automated iterative scheme was employed to obtain the self-consistent $U_{sc}$ parameter while simultaneously optimizing the structure and desired spin state: starting from an empirical $U$ of 4.3 eV, the energies of all possible occupation matrices for a spin state were computed. There are five possible occupation matrices corresponding to the HS state of ferrous iron with $3d^6$ configuration (S=2), while there are ten possibilities for the LS state (S=0). The electronic configuration, i.e., occupation matrix, with the lowest energy, was selected for further structural optimization of lattice parameters and atomic positions. Then a new $U$ parameter is recalculated for further structural optimization. The process continued until mutual convergence of structure and $U$ is achieved for a convergence threshold of 0.01 eV for the $U$ parameter and the convergence criteria mentioned above for structural optimizations. Only the lowest energy configuration was adopted in subsequent calculations. Finite temperature effects on the static DFT energy were included using the Mermin functional with the Fermi-Dirac smearing \cite{23,24}. The temperature-dependent electronic entropy was obtained from 0 K to 4500 K every 500 K and then interpolated. The scheme used here was described in Ref.\cite{25}. We also computed the excess free energy from non-ideal mixing of HS and LS states which is described together with the ideal solution model in Sec. V.

With large unit cells containing 64 atoms, phonon calculations were performed using the finite-displacement method, using Phonopy code \cite{26} with force constants computed by Quantum ESPRESSO. Vibrational density of states (VDOSs) were obtained using a $q$-point $20\times20\times20$ mesh. The vibrational contribution to the free energy was calculated using the quasiharmonic approximation \cite{27} with the $qha$ code \cite{28}.

\section{Electronic structure fp with $x_{Fe}$=0.03125$\%$ (fp3)}

To first have a clear picture of the electronic structure of fp, we consider only one Mg substitution by Fe in the 64-atom supercell, i.e., FeMg$_{31}$O$_{32}$, $x_{Fe}$=0.03125 (fp3 hereafter), as shown in Fig. 1(a). In this case, there is no Fe-Fe interaction so that the energy levels of the 3$d$ orbitals in the ferrous Fe can be well-identified. Ferrous Fe(Fe$^{2+}$)   with 3$d^6$ electronic configurations has six O neighbors in octahedral coordination. The octahedral crystal field splits the fivefold d-orbital degeneracy, producing a doublet with e$_g$ symmetry and a triplet with t$_{2g}$ symmetry. Because the t$_{2g}$ orbitals are pointing away from the oxygen neighbors, the t$_{2g}$ orbitals have lower energy than the e$_g$ orbitals, shown in Fig 1(c). In the HS state at low pressures, following Hund's rule, five of six electrons occupy five spin-up orbitals, and the remaining minority electron fills one of the t$_{2g}$ orbitals, as shown in Fig.1(a). The Fe-O octahedron is Jahn-Teller (JT) distorted in this electronic configuration, i.e., it has two short and four long bonds. At P=0 GPa, the difference between the short and long bonds is 2.4$\%$ in Fig. 1(a). The JT distortion, in turn, causes further energy splitting within the $e_g$ and t$_{2g}$ levels so that one can see a slight energy difference between $d_{z^2}$ and $d_{x^2-y^2}$, as well as $d_{xy}$ and $d_{zx}$ ($d_{zy}$). Ferrous Fe can also exhibit the low-spin (LS) state with all six electrons in the t$_{2g}$ orbitals, as shown in Fig. 1(d). Because the occupied $t_{2g}$ orbitals have cubic symmetry, as shown in Fig. 1(b), there is no JT distortion in the equilibrated LS state. The volume of the FeO octahedron in the LS state is smaller than that in the HS state. Comparing the lattices at 0 GPa in Fig. 1(a) and (b), the octahedron volume of the LS is 7.4$\%$ smaller than the one with Fe$^{2+}$ in the HS state. Both HS and LS of fp3 are insulators.

\begin{figure}[t]
\includegraphics[width=0.47\textwidth]{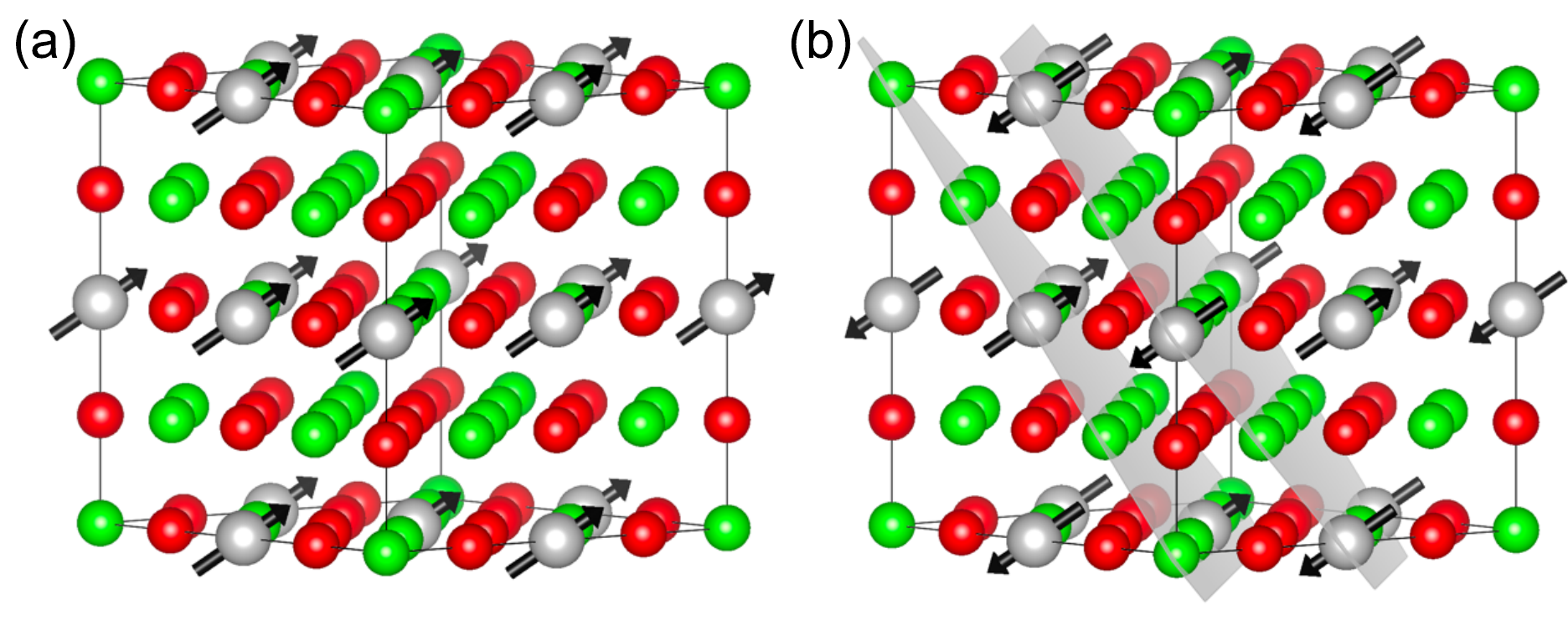}
\caption{\label{fig:fig2}(a) Supercell structure of fp18 with (a) ferromagnetic and (b) antiferromagnetic configurations. The [111] planes are indicated in (b). However, in our collinear spin calculations, the direction of the spin magnetic moment is not relevant.}
\end{figure}

\section{Spin crossover of fp18 at T=0K}
We now perform static calculations to study the spin crossover in fp at T=0K. For a pyrolitic mantle composition, the fp volume fraction should be around 0.15-0.20 \cite{a1} . Here we focus on the spin crossover in the Fe$_6$Mg$_{26}$O$_{32}$ lattice, i.e., $x_{Fe}$=0.1875 (fp18 hereafter), while some results on fp3 are also included for comparison. To construct the supercell structure, we distribute 6 Fe uniformly by occupying the face centers and edge centers, as shown in Fig. 2. Because fp is a solid solution of FeO and MgO, a uniform distribution is more relevant to the real situation. Moreover, it has been shown that different types of Fe configurations have only a small effect on the spin crossover \cite{29}. With $x_{Fe}$=0.1875, one would expect the exchange interaction between Fe ions in the HS state to be sizable because of the small Fe-Fe distance. Here we consider both the ferromagnetic (FM) and antiferromagnetic (AFM) configurations with spins aligned as in Figs. 2(a,b). Considering the AFM spin configuration in FeO-B1 \cite{16}, we assign opposite magnetic moments in the AFM configuration alternating in neighboring [111] planes.

\begin{figure}[t]
\includegraphics[width=0.48\textwidth]{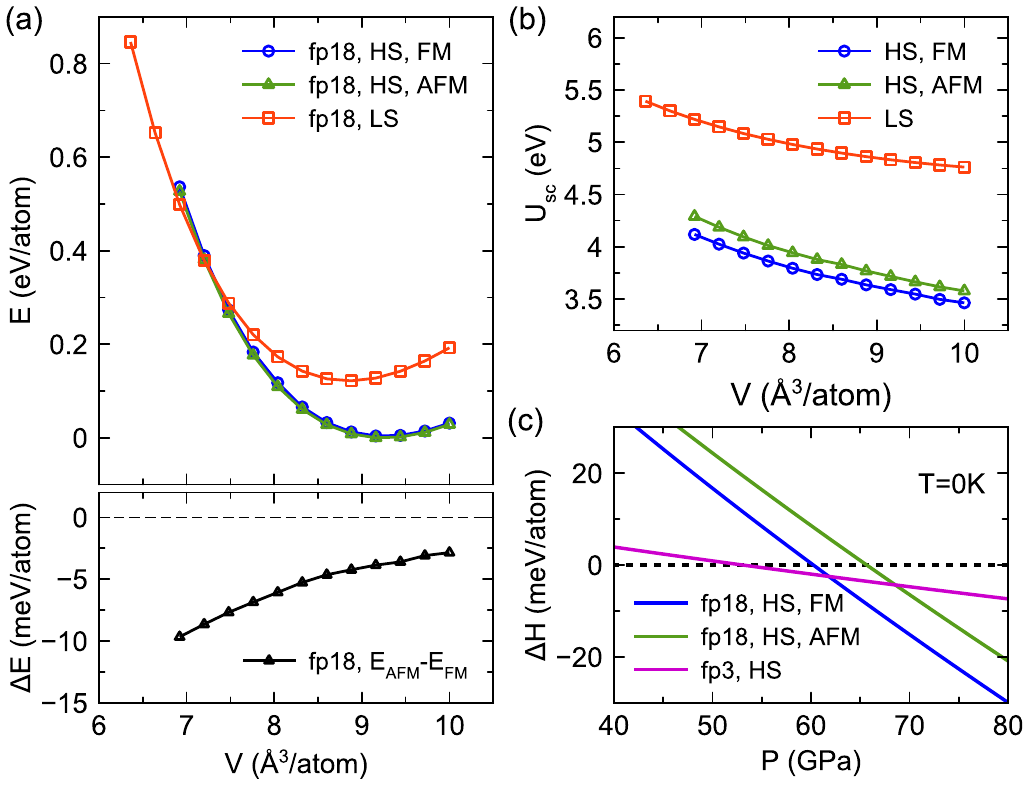}
\caption{\label{fig:fig3}(a) Upper panel shows the E vs. V curves for the three states of fp18. The lower panel shows the energy difference between the HS-AFM and HS-FM. (b) The self-consistent Hubbard parameters U for different spin states. (c) The enthalpy difference between HS and LS ($\Delta H=H_{HS}-H_{LS}$).}
\end{figure}

Figure 3 (a) shows the volume-dependent energy for fp18 HS states with FM and AFM configurations and LS state. The HS states have lower energy at large volumes (low or negative pressures) than the LS state. The AFM configuration has the lowest energy of all. Therefore, at ambient conditions, the ground state is the HS with AFM magnetic order. This is consistent with the experimental measurement that AFM is the ground state below the Néel temperature [9]. With decreasing volume, the LS state energy decreases w.r.t. that of the HS states. This is mainly because the energy splitting between t$_{2g}$ and e$_g$ increases with increasing pressure, leading to the spin crossover. Figure 3(b) shows the self-consistent Hubbard parameters for the different spin states. The $U_{sc}$ values of the LS state are systematically higher than the HS states regardless of volume. This trend is similar to fp3 and previous studies of Hubbard parameters of HS and LS states in the FeO system [16]. The self-consistent $U$ value of HS-FM and HS-AFM also shows a slight difference. The energy-volume data are fitted by the third-order Birch-Murnaghan (BM) equation of state (EoS) using the least squares method. The enthalpies are obtained from the fitted BM-EoS and are shown in Fig. 3(c). Based on the enthalpy difference, the transition pressure from HS to LS can be identified. In fp18, the transition pressures are 60 GPa for the FM state and 66 GPa for the AFM state. By performing similar calculations with fp3, we find the HS-LS transition pressure in fp3 is 53 GPa. Therefore the spin crossover pressure increases with an increasing iron concentration in ferropericlase. This transition pressure in fp18 is in good agreement with the experimental measurement of ferropericlase with $x_{Fe}$=0.17 at room temperature [7], which is $\sim$60GPa. A change in the distribution of iron in the supercell of fp18 changes the transition pressure by a few GPa only [29], but it gives rise to a distribution of transition pressures. 

\begin{figure}[t]
\includegraphics[width=0.4\textwidth]{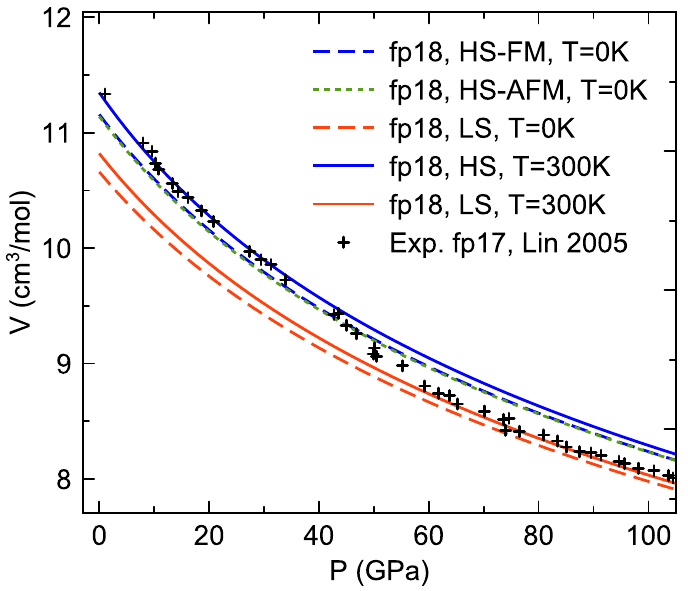}
\caption{\label{fig:fig4}Compression curve of fp18. The experimental data is at 300K with $x_{Fe}$=17$\%$ ferropericlase \cite{7}.}
\end{figure}

The fp18 compression curves obtained from the zero-kelvin EoS are shown in Fig. 4. The curves of HS-FM and HS-AFM states are almost overlapped here. The experimental measurement at room temperature with $x_{Fe}$=0.17 ferropericlase \cite{7} is also included for comparison. The calculated zero-kelvin compression curves are close but systematically smaller than the experimental data. This result is reasonable because no temperature effect is included yet. As discussed later, vibrational effects at finite temperature further improve the agreement with experiments.

Figure 5 shows the projected density of states (DOS) of Fe-3$d$ for fp18. Both HS and LS DOS are qualitatively similar to that of fp3 in Figure 1. However, a larger Fe concentration caused stronger crystal field splitting, leading to complicated energy levels of different orbitals. Nevertheless, both HS and LS states remain to be the insulating states. The gaps are around 2 eV and almost independent of the pressure increasing.

\begin{figure}[b]
\includegraphics[width=0.47\textwidth]{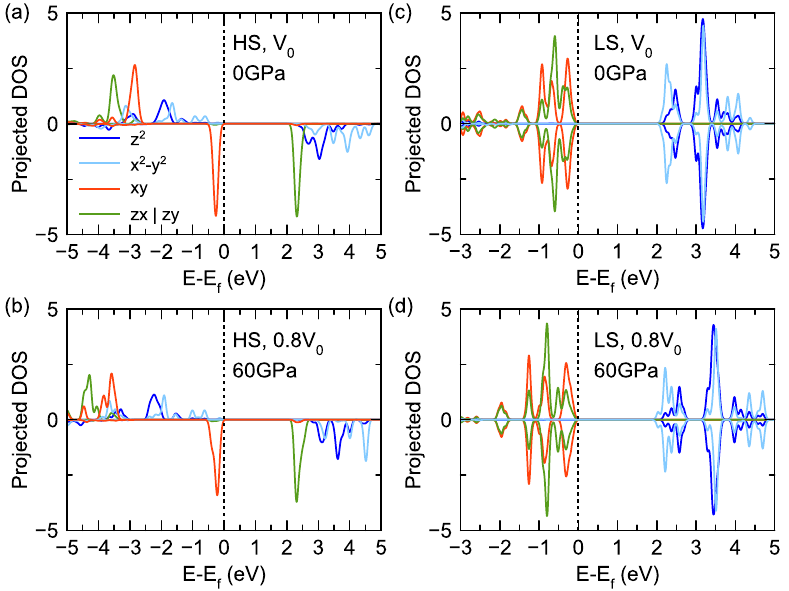}
\caption{\label{fig:fig5}Projected density of state (DOS) of the Fe 3d orbitals in fp18 with (a) HS at 0 GPa; (b) HS at 60 GPa; (c) LS at 0 GPa; (d) LS at 60 GPa.}
\end{figure}

\section{Finite temperature effect on the spin crossover}
Phonon calculations are performed with LDA+$U_{sc}$ for all calculated volumes of both HS- and LS-fp18. Figure 6 shows examples of three vibrational densities of state (VDOS) from low to high pressures. All other VDOS are shown in Supplementary Fig. S1. With increasing pressure, the phonon frequencies are shifted towards higher energies. No imaginary frequency is found in either HS or LS state up to 100 GPa. This is consistent with the recent phonon calculations in $x_{Fe}$=0.0625 fp with density functional perturbation theory \cite{29}. Therefore, both HS and LS states of fp with iron concentrations lower than 0.1875 are dynamically stable. 

\begin{figure}[t]
\includegraphics[width=0.5\textwidth]{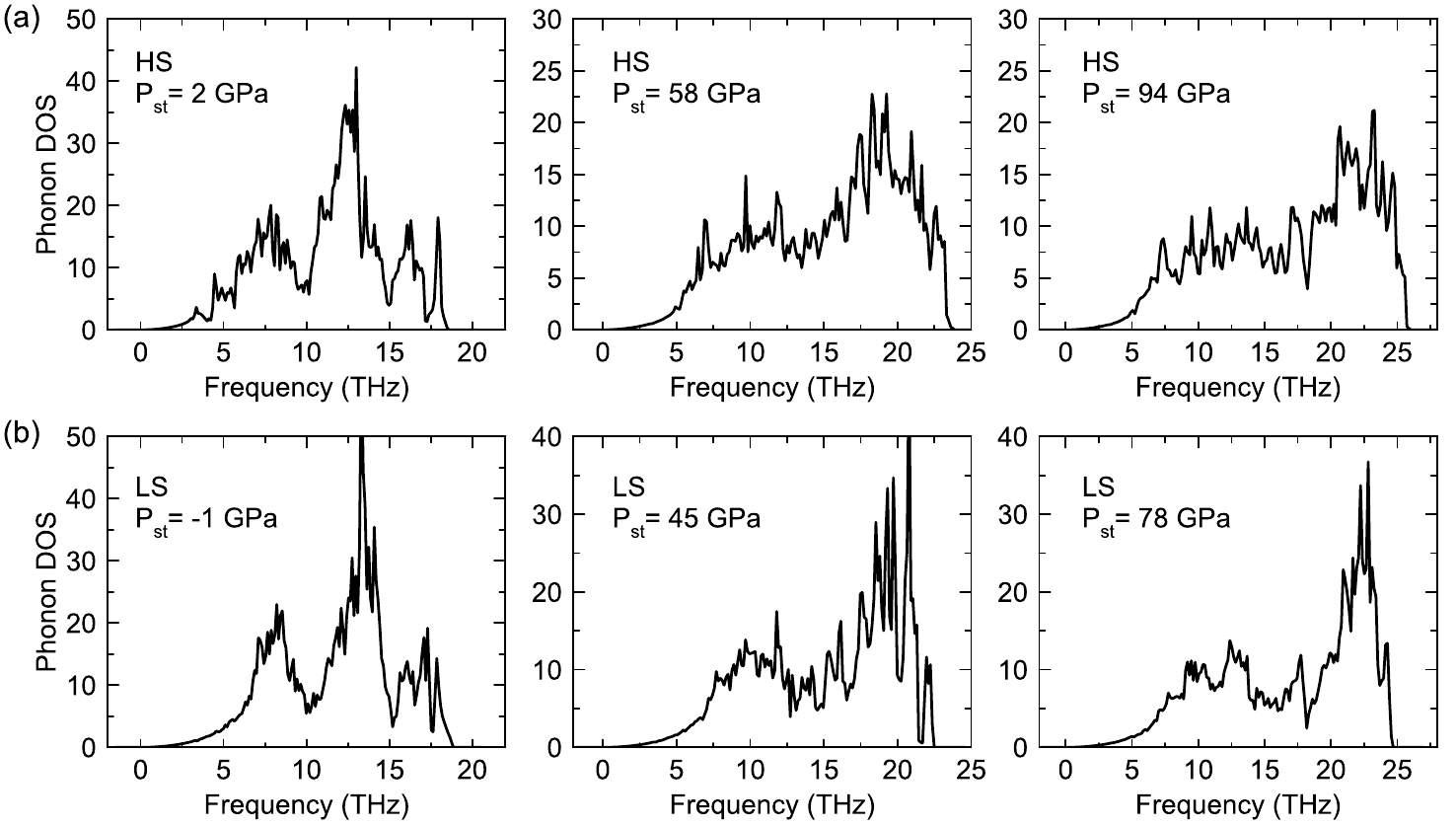}
\caption{\label{fig:fig6}Vibrational DOS for (a) HS and (b) LS fp18 with three volumes. The upper and lower panels correspond to the same volume. Their static pressures at zero kelvin are indicated in the figure.}
\end{figure}

\begin{figure}[b]
\includegraphics[width=0.45\textwidth]{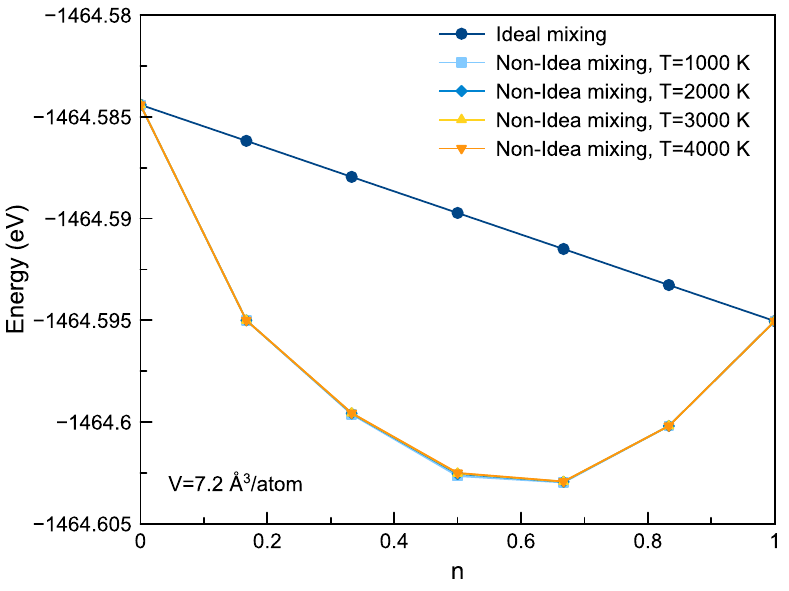}
\caption{\label{fig:fig6}Static energy as a function n at constant volume V=7.2$\AA^3$/atom for ideal mixing and non-ideal mixing models.}
\end{figure}

With the inclusion of vibrational entropy and electronic entropy described by the Mermin functional in static free energy calculations \cite{23,24}, quasiharmonic calculations are performed to compute the free energy and EoS at finite temperatures. With the inclusion of thermal electronic excitation effects on the static free energy, the compression curve of both HS and LS at 300K agrees very well with room-temperature experimental measurements \cite{7}, as shown in Fig. 4. To obtain the HS-LS phase boundary at finite temperature, we further consider the non-ideal mixing of HS and LS states and its contribution to the free energy. Using $n$ to represent the fraction of LS states, the total free energy of an ideal solid solution of HS and LS can be written as

\begin{widetext}

\begin{equation}
G_{ideal}\left(P,T,n\right)=(1-n)G_{HS}(P,T)+nG_{LS}(P,T)+G_{mix}^{ideal}(n),
\end{equation}

\end{widetext}

where $G_{HS/LS}$ is the molar Gibbs free energy of the pure HS/LS states, i.e.,

\begin{equation}
G_{HS/LS}\left(P,T\right)=G_{HS/LS}^{stat+vib}\left(P,T\right)+G_{HS/LS}^{mag},
\end{equation}

where $G_{HS/LS}^{stat+vib}(P,T)$ is the Gibbs free energy containing static and vibrational contribution and $G_{HS/LS}^{mag}$ is the magnetic contribution. $G_{HS/LS}^{mag}$ is a purely entropic contribution that one can estimate approximately as

\begin{equation}
G^{mag}=-k_BTx_{Fe}ln\left[m\left(2S+1\right)\right],
\end{equation}

where $S$ and $m$ are the spin and electronic configuration (orbital) degeneracies of iron. In HS, S=2 and m=3. In LS, S=0 and m=1. The ideal free energy of mixing is given by the mixing entropy as

\begin{widetext}

\begin{equation}
G_{mix}^{ideal}(n)=-TS_{mix}=-k_B{Tx}_{Fe}[n\ln{n}+\left(1-n\right)\ln{\left(1-n\right)}].
\end{equation}

\end{widetext}

Eq. (4) gives the free energy of mixing of the ideal solution of HS and LS states, where $n$ is the LS fraction. By minimizing the free energy in Eq. (1) with respect to $n$, one obtains

\begin{equation}
n_{ideal}=\frac{1}{1+m(2S+1)exp[\frac{\Delta G_{LS-HS}(P,T)}{k_BTx_{Fe}}]},
\end{equation}
where ${\Delta G}_{LS-HS}(P,T)=G_{LS}\left(P,T\right)-G_{HS}\left(P,T\right)$. 

We now examine the effect of non-ideal mixing of HS and LS states. In this case, different HS/LS configurations are considered for a single Mg/Fe atomic arrangement in fp18 and several $n$ values. The single Mg/Fe configuration in the (Mg$_{26}$Fe$_6$)O$_{32}$ supercell is sampled for $n=\frac{1}{6}$,$\frac{2}{6}$,$\frac{3}{6}$,$\frac{4}{6}$ and $\frac{5}{6}$. These configurations are listed in Supplementary Table S1 and Supplementary Figure S2. The static energy $\varepsilon_i$ of the $i^{th}$ non-equivalent atomic configuration is computed in a large pressure range, using the consistent Hubbard parameters of fp18 shown in Fig. 3(b). The non-ideal mixing energy can be obtained by making a Boltzmann ensemble average for all non-equivalent arrangements. Then

\begin{equation}
E_{non-ideal}=\sum_{i}{{g_ip}_i\varepsilon_i},
\end{equation}

where $g_i$ is the multiplicity of the $i^{th}$ non-equivalent atomic configuration and $p_i$ is the Boltzmann factor as $p_i=\frac{e^{\varepsilon_i/k_BT}}{\sum_{i}{g_ie^{\varepsilon_i/k_BT}}}$. Figure 7 shows that the results from non-ideal mixing deviate from the one in the ideal mixing model, indicating that the non-ideal mixing effect is relatively significant when the HS and LS states have similar static energies at the same volume. . The excess energy can be obtained by calculating the energy difference between ideal and non-ideal models as $E_{ex}\left(n\right)=E_{non-ideal}\left(n\right)-E_{ideal}\left(n\right)$.  As can be seen in Fig. 7, the temperature dependence of $E_{ex}\left(n\right)$ is insignificant. We then include an excess free energy term, $G_{ex}(P,T,n)$, of non-ideal mixing in Eqn. (1),

\begin{widetext}
\begin{equation}
G_{non-ideal}\left(P,T,n\right)=G_{ideal}\left(P,T,n\right)+G_{ex}(P,T,n).
\end{equation}
\end{widetext}

Here we assume the excess free energy is mainly contributed by the static part so that we keep the vibrational contributions the same as the one in the ideal mixing model, i.e. $G_{ex}\left(P,T,n\right)\approx H_{ex}^{stat}\left(P,T,n\right)$, where the T-dependence is negligible as in $E_{ex}\left(n\right)$. The excess enthalpy can be obtained by fitting $E_{non-ideal}$ with $3^{rd}$ BM-EOS, obtaining pressure and adding the $P_{non-ideal}V$ term. Similar to solving the ideal mixing model, $n_{non-ideal}$ can be obtained by minimizing Eqn. (7) with respect to $n$, which leads to

\begin{widetext}

\begin{equation}
f\left(P,T,n\right)={\Delta G}_{LS-HS}\left(P,T\right)+\frac{\partial H_{ex}(P,T,n)}{\partial n}+k_BTx_{Fe}\ln{\left[\frac{n}{1-n}\left(m\left(2S+1\right)\right)\right]}=0.
\end{equation}

\end{widetext}

We $numerically$ solve Eqn. (8) for $n_{non-ideal}$ by first fitting $H_{ex}\left(n\right)$ with polynomial functions at each pressure and temperature, as shown in Fig. S2. The obtained $n_{non-ideal}(P,T)$ are plotted as a function of pressure and temperature in Fig. 8. 

\begin{figure}[t]
\includegraphics[width=0.5\textwidth]{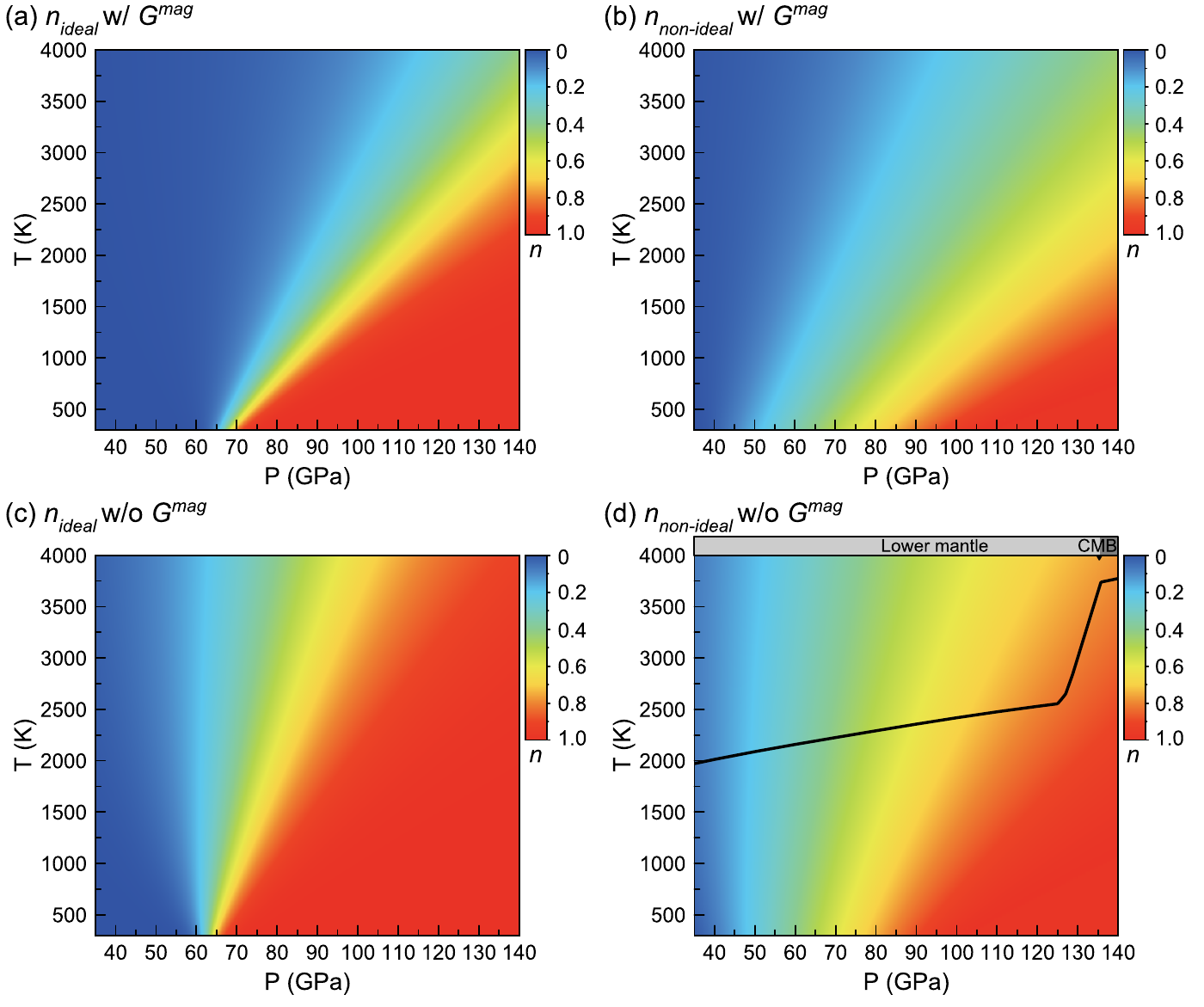}
\caption{\label{fig:fig8}Temperature-dependent spin-crossover ranges with fp18 based on the different models. (a) and (b) are ideal and non-ideal mixing models with $G^{mag}$; (c) and (d) are ideal and non-ideal mixing models without $G^{mag}$. The black line indicates the geotherm \cite{30,31}.}
\end{figure}

We now analyze how the magnetic entropy contribution and non-ideal mixing affect the temperature dependence of the HS-LS spin-crossover pressures in Fig. 8. Comparing the results from ideal and non-ideal mixing models (Fig. 8 (a) vs. (b), or (c) vs. (d)), the HS-LS mixing range is much broader in the non-ideal mixing models. At room temperature, the crossover pressure range is $\sim$ 5GPa in the ideal mixing model, while it is $\sim$30GPa in the non-ideal mixing model. This MS pressure range broadens further at higher temperatures. In principle, the non-ideal mixing model should be closer to the real situation than the ideal mixing model. The wide crossover range agrees better with the experimental data \cite{6}. By comparing Fig. 8 (a) and (c) (or (b) and (d)), we find $G^{mag}$ significantly increases the Clapeyron slope of the spin crossover range. The slope of the phase boundary in Fig. 8(d) is more similar to the previous experimental data from Ref. \cite{6} than others. This might occur because the current $G^{mag}$ is an approximate analytical estimate of the largest possible contribution of the magnetic entropy. In reality, the local spin magnetic moment at high temperatures should not be as large as S = 2 for the HS state. Therefore, this magnetic entropic effect is likely overestimated. Taking these factors into account, Fig. 8(d) might represent the most realistic situation for the HS-LS spin cross in fp18. In Fig. 8(d), the spin crossover starts at $\sim$ 45 GPa. High temperatures do not significantly affect the HS fraction at $\sim$45 GPa, which agrees with the experimental data \cite{6}. We include the mantle geotherm for a pyrolytic composition \cite{30} in Figs. 8 (d). The spin-crossover starts at $\sim$45 GPa, 2100 K and ends at $\sim$115 GPa, 2400 K along the geotherm. These results indicate the MS state should predominate in fp in most of the lower mantle. fp near the core-mantle boundary (CMB) should be mainly in the LS state. We note that thermal electronic excitation effects described using the Mermin functional do not significantly affect the phase boundary, as shown in Supplementary Fig. S4. This is expected as both HS and LS are insulators.

\section{Summary}
In summary, we revisited the HS-LS crossover in fp by performing LDA+$U_{sc}$ calculations. The Hubbard parameters $U$ are determined self-consistently ($U_{sc}$) using density functional perturbation theory (DFPT). The $U_{sc}$ parameter depends on pressure, spin state, electronic and atomic configuration, etc., and varies by 1-2 eV. The AFM configuration is found to be the ground state at low temperatures, consistent with experiments. The energy difference between FM and AFM configurations is less than 10meV/atom at T=0K, and magnetic ordering has a relatively minor impact on the spin crossover pressure in static calculations for $x_{Fe}$ = 0.1875. Phonon spectra are computed for both HS and LS states. No imaginary frequencies are found in any case from 0 GPa to 120 GPa, confirming phonon stability in fp in the entire pressure range of the Earth's mantle. Quasiharmonic free energy calculations offer \textit{ab initio} compression curves for HS and LS states in good agreement with experimental data at room temperature. The HS-LS phase diagram is obtained by including all finite-temperature effects, i.e., vibrational, magnetic, electronic, and non-ideal HS-LS mixing contributions. The non-ideal HS-LS solid-solution mixing model gives a crossover starting at $\sim$ 45 GPa at room temperature and is considerably broader than previous calculations. The magnetic entropy is found to affect the Clapeyron slope of the HS-LS crossover significantly. Considering these effects, the mixed spin state is predicted to predominate the fp throughout most of the lower mantle.

\begin{acknowledgments}
This work was supported primarily by National Science Foundation awards EAR-1918126 and EAR-1918134. We acknowledge partial support from the U.S. Department of Energy Grant DE-SC0019759. We also acknowledge the computer resources from the Extreme Science and Engineering Discovery Environment (XSEDE), which is supported by the National Science Foundation grant number ACI-1548562.
\end{acknowledgments}

\bibliographystyle{apsrev4-2}

\begin{thebibliography}{0}%
\makeatletter
\providecommand \@ifxundefined [1]{%
 \@ifx{#1\undefined}
}%
\providecommand \@ifnum [1]{%
 \ifnum #1\expandafter \@firstoftwo
 \else \expandafter \@secondoftwo
 \fi
}%
\providecommand \@ifx [1]{%
 \ifx #1\expandafter \@firstoftwo
 \else \expandafter \@secondoftwo
 \fi
}%
\providecommand \natexlab [1]{#1}%
\providecommand \enquote  [1]{``#1''}%
\providecommand \bibnamefont  [1]{#1}%
\providecommand \bibfnamefont [1]{#1}%
\providecommand \citenamefont [1]{#1}%
\providecommand \href@noop [0]{\@secondoftwo}%
\providecommand \href [0]{\begingroup \@sanitize@url \@href}%
\providecommand \@href[1]{\@@startlink{#1}\@@href}%
\providecommand \@@href[1]{\endgroup#1\@@endlink}%
\providecommand \@sanitize@url [0]{\catcode `\\12\catcode `\$12\catcode
  `\&12\catcode `\#12\catcode `\^12\catcode `\_12\catcode `\%12\relax}%
\providecommand \@@startlink[1]{}%
\providecommand \@@endlink[0]{}%
\providecommand \url  [0]{\begingroup\@sanitize@url \@url }%
\providecommand \@url [1]{\endgroup\@href {#1}{\urlprefix }}%
\providecommand \urlprefix  [0]{URL }%
\providecommand \Eprint [0]{\href }%
\providecommand \doibase [0]{https://doi.org/}%
\providecommand \selectlanguage [0]{\@gobble}%
\providecommand \bibinfo  [0]{\@secondoftwo}%
\providecommand \bibfield  [0]{\@secondoftwo}%
\providecommand \translation [1]{[#1]}%
\providecommand \BibitemOpen [0]{}%
\providecommand \bibitemStop [0]{}%
\providecommand \bibitemNoStop [0]{.\EOS\space}%
\providecommand \EOS [0]{\spacefactor3000\relax}%
\providecommand \BibitemShut  [1]{\csname bibitem#1\endcsname}%
\let\auto@bib@innerbib\@empty
\end{thebibliography}%


\begin{thebibliography}{50}
\bibitem{a1}      Wentzcovitch R M, Yu Y G and Wu Z 2010 Thermodynamic properties and phase relations in mantle minerals investigated by first principles quasiharmonic theory Rev. Mineral. Geochemistry 71 59–98
\bibitem{1} 	 Tsuchiya T, Wentzcovitch R M, da Silva C R S and de Gironcoli S 2006 Spin Transition in Magnesiowüstite in Earth's Lower Mantle Phys. Rev. Lett. 96 198501
\bibitem{2}	 Wu Z, Justo J F, da Silva C R S, de Gironcoli S and Wentzcovitch R M 2009 Anomalous thermodynamic properties in ferropericlase throughout its spin crossover Phys. Rev. B 80 014409
\bibitem{3}	 Wentzcovitch R M, Justo J F, Wu Z, da Silva C R S, Yuen D A and Kohlstedt D 2009 Anomalous compressibility of ferropericlase throughout the iron spin cross-over Proc. Natl. Acad. Sci. 106 8447–52
\bibitem{4}	 Wu Z and Wentzcovitch R M 2014 Spin crossover in ferropericlase and velocity heterogeneities in the lower mantle Proc. Natl. Acad. Sci. 111 10468–72
\bibitem{5}	 Badro J, Fiquet G, Guyot F, Rueff J P, Struzhkin V V., Vankó G and Monaco G 2003 Iron partitioning in Earth's mantle: Toward a deep lower mantle discontinuity Science 300 789–91
\bibitem{6}	 Lin J-F, Vanko G, Jacobsen S D, Iota V, Struzhkin V V., Prakapenka V B, Kuznetsov A and Yoo C-S 2007 Spin Transition Zone in Earth's Lower Mantle Science 317 1740–3
\bibitem{7}	 Lin J F, Struzhkin V V., Jacobsen S D, Hu M Y, Chow P, Kung J, Liu H, Mao H K and Hemley R J 2005 Spin transition of iron in magnesiowüstite in the Earth's lower mantle Nature 436 377–80
\bibitem{8}	 Fei Y, Zhang L, Corgne A, Watson H, Ricolleau A, Meng Y and Prakapenka V 2007 Spin transition and equations of state of (Mg, Fe)O solid solutions Geophys. Res. Lett. 34 1–5
\bibitem{9}	 Lyubutin I S, Struzhkin V V., Mironovich A A, Gavriliuk A G, Naumov P G, Lin J F, Ovchinnikov S G, Sinogeikin S, Chow P, Xiao Y and Hemley R J 2013 Quantum critical point and spin fluctuations in lower-mantle ferropericlase Proc. Natl. Acad. Sci. U. S. A. 110 7142–7
\bibitem{10}	 Mao Z, Lin J F, Liu J and Prakapenka V B 2011 Thermal equation of state of lower-mantle ferropericlase across the spin crossover Geophys. Res. Lett. 38 2–5
\bibitem{11}	 Cococcioni M and de Gironcoli S 2005 Linear response approach to the calculation of the effective interaction parameters in the LDA+U method Phys. Rev. B 71 035105
\bibitem{12}	 Kulik H J, Cococcioni M, Scherlis D A and Marzari N 2006 Density Functional Theory in Transition-Metal Chemistry: A Self-Consistent Hubbard U Approach Phys. Rev. Lett. 97 103001
\bibitem{13}	 Hsu H, Umemoto K, Wu Z and Wentzcovitch R M 2010 Spin-state crossover of iron in lower-mantle minerals: Results of DFT+U investigations Rev. Mineral. Geochemistry 71 169–99
\bibitem{14}	 Cococcioni M and Marzari N 2019 Energetics and cathode voltages of LiMPO4 olivines (M=Fe , Mn) from extended Hubbard functionals Phys. Rev. Mater. 3 033801
\bibitem{15}	 Floris A, Timrov I, Himmetoglu B, Marzari N, De Gironcoli S and Cococcioni M 2020 Hubbard-corrected density functional perturbation theory with ultrasoft pseudopotentials Phys. Rev. B 101 064305
\bibitem{16}	 Sun Y, Cococcioni M and Wentzcovitch R M 2020 LDA+U calculations of phase relations in FeO Phys. Rev. Mater. 4 063605
\bibitem{17}	 Timrov I, Marzari N and Cococcioni M 2018 Hubbard parameters from density-functional perturbation theory Phys. Rev. B 98 085127
\bibitem{18}	 Dudarev S L, Botton G A, Savrasov S Y, Humphreys C J and Sutton A P 1998 Electron-energy-loss spectra and the structural stability of nickel oxide: An LSDA+U study Phys. Rev. B 57 1505–9
\bibitem{19}	 Giannozzi P, Baroni S, Bonini N, Calandra M, Car R, Cavazzoni C, Ceresoli D, Chiarotti G L, Cococcioni M, Dabo I, Dal Corso A, De Gironcoli S, Fabris S, Fratesi G, Gebauer R, Gerstmann U, Gougoussis C, Kokalj A, Lazzeri M, Martin-Samos L, Marzari N, Mauri F, Mazzarello R, Paolini S, Pasquarello A, Paulatto L, Sbraccia C, Scandolo S, Sclauzero G, Seitsonen A P, Smogunov A, Umari P and Wentzcovitch R M 2009 QUANTUM ESPRESSO: a modular and open-source software project for quantum simulations of materials J. Phys. Condens. Matter 21 395502
\bibitem{20}	 Giannozzi P, Andreussi O, Brumme T, Bunau O, Buongiorno Nardelli M, Calandra M, Car R, Cavazzoni C, Ceresoli D, Cococcioni M, Colonna N, Carnimeo I, Dal Corso A, De Gironcoli S, Delugas P, Distasio R A, Ferretti A, Floris A, Fratesi G, Fugallo G, Gebauer R, Gerstmann U, Giustino F, Gorni T, Jia J, Kawamura M, Ko H Y, Kokalj A, Kücükbenli E, Lazzeri M, Marsili M, Marzari N, Mauri F, Nguyen N L, Nguyen H V., Otero-De-La-Roza A, Paulatto L, Poncé S, Rocca D, Sabatini R, Santra B, Schlipf M, Seitsonen A P, Smogunov A, Timrov I, Thonhauser T, Umari P, Vast N, Wu X and Baroni S 2017 Advanced capabilities for materials modelling with Quantum ESPRESSO J. Phys. Condens. Matter 29 465901
\bibitem{21}	 Dal Corso A 2014 Pseudopotentials periodic table: From H to Pu Comput. Mater. Sci. 95 337–50
\bibitem{22}	 Anisimov V I, Zaanen J and Andersen O K 1991 Band theory and Mott insulators: Hubbard U instead of Stoner I Phys. Rev. B 44 943–54
\bibitem{23}	 Mermin N D 1965 Thermal properties of the inhomogeneous electron gas Phys. Rev. 137 A1441
\bibitem{24}	 Wentzcovitch R M, Martins J L and Allen P B 1992 Energy versus free-energy conservation in first-principles molecular dynamics Phys. Rev. B 45 11372–4
\bibitem{25}	 Zhuang J, Wang H, Zhang Q and Wentzcovitch R M 2021 Thermodynamic properties of $\epsilon$-Fe with thermal electronic excitation effects on vibrational spectra Phys. Rev. B 103 144102
\bibitem{26}	 Togo A and Tanaka I 2015 First principles phonon calculations in materials science Scr. Mater. 108 1–5
\bibitem{27}	 Wallace D C 1972 Thermodynamics of Crystals (Mineola: Dover)
\bibitem{28}	 Qin T, Zhang Q, Wentzcovitch R M and Umemoto K 2019 qha: A Python package for quasiharmonic free energy calculation for multi-configuration systems Comput. Phys. Commun. 237 199–207
\bibitem{29}	 Marcondes M L, Zheng F and Wentzcovitch R M 2020 Phonon dispersions throughout the iron spin crossover in ferropericlase Phys. Rev. B 102 104112
\bibitem{30}	 Valencia-Cardona J J, Shukla G, Wu Z, Houser C, Yuen D A and Wentzcovitch R M 2017 Influence of the iron spin crossover in ferropericlase on the lower mantle geotherm Geophys. Res. Lett. 44 4863–71
\bibitem{31}	 Stacey F D and Davis P M 2008 Physics of the Earth vol 193 (Cambridge: Cambridge University Press)

\end{thebibliography}


\end{document}